\documentclass{sig-alternate}
\usepackage{url}
\usepackage{multirow}

\usepackage{times}
\usepackage{latexsym}
\usepackage{amsmath}
\usepackage{multirow}
\usepackage{graphicx}
\usepackage{multirow}
\usepackage{subfigure}
\usepackage{url}
\usepackage{comment}

\usepackage{color}
\usepackage{url}
\usepackage{comment}
\usepackage{enumitem}
  {\begin{itemize}[topsep=0pt, partopsep=0pt] \footnotesize%
    \setlength{\itemsep}{0pt}%
    \setlength{\parskip}{0pt}%
    }%
  {\end{itemize}}

  {\begin{itemize}[topsep=10pt, partopsep=0pt] \footnotesize%
    \setlength{\itemsep}{2pt}%
    \setlength{\parskip}{0pt}%
    }%
  {\end{itemize}}

\begin{document}

\title{What a Nasty day: \\Exploring Mood-Weather Relationship from Twitter}

\numberofauthors{3} 
\author{
\alignauthor
Jiwei Li\\
       \affaddr{Department of Computer Science}\\
      \affaddr{Stanford University}\\
       \affaddr{Stanford, CA 94305, USA}\\
       \affaddr{jiweil@stanford.edu}
\alignauthor
Xun Wang\\
       \affaddr{Department of Computer Science}\\
       \affaddr{Peking University}\\
       \affaddr{Beijing, 100871, P.R. China}\\
       \affaddr{xun$\_$wang@gmail.com}
\alignauthor       
Eduard Hovy\\
	  \affaddr{Language Technology Institute}\\
\affaddr{School of Computer Science}\\
	  \affaddr{Carnegie Mellon University}\\
	  \affaddr{Pittsburgh, PA 15213, USA}\\
	   \affaddr{hovy@cmu.edu}
}
\maketitle

\begin{abstract}
While it has long been believed in psychology that weather somehow influences human's mood, the debates have been going on for decades about how they are correlated.
In this paper, we try to study this long-lasting topic by harnessing a new source of data compared from traditional 
psychological researches: Twitter.
We analyze 2 years' twitter data collected by twitter API which amounts to
$10\%$ of all postings and try to reveal the correlations between multiple dimensional structure of human mood with meteorological effects. 
Some of our findings confirm existing hypotheses, while others contradict them. We are hopeful that our approach, along with the new data source, 
can shed on the long-going debates on weather-mood correlation. 
\end{abstract}


\section{Introduction}
There is a general agreement that weather somehow affects individual's emotions, especially 
considering strong quantitative evidence from biologists that
happiness related chemicals rose or dip in brain as humans are exposed to different types of weather conditions (See Section 2).
Regrettably, conclusions from psychologist are amazingly diverse, or even opposite (See Section 2),
leaving
the debate of mood-weather correlation going on for decades.
Traditional surveys in psychological literature mostly rely on questionnaire, pulls or course surveys, which are not only time-consuming but expensive, disabling large-scale tracing surveys. 
Conclusions drawn from the limited data set are extremely sensitive to individuals in the survey, leading to diverse or even opposite conclusions.

\begin{figure}
\centering
\includegraphics[width=3.5in]{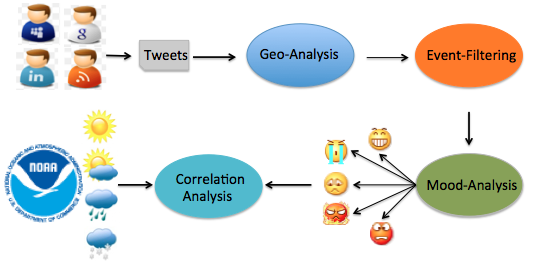}
\caption{Pipeline for mood-weather relationship analysis based on Twitter.}
\label{fig:1}
\end{figure}

The drastic popularity of online social media, provides a much faster, more comprehensive but far less expensive alternative to traditional pulls. 
Users regularly post tweets broadcasting their moods, thoughts, ideas and states encountered in their lives.
Their user-generated content provides direct or indirect evidence to their mood or emotions. 
Twitter data exhibits an important property that surpasses questionnaire-based researches: the multitude messages from great numbers of users in a long-time span.
Such property enables {\it aggregated analysis} \cite{hopkins2010method,o2010tweets}, where 
data errors or the influence from nuisance factors (i.e., personality traits, gender, or age)
can be canceled out by the great amount of population in study.
As the flexibility, robustness and sensitivity of Twitter mood indicators have been demonstrated in many existing researches \cite{dodds2010measuring,dodds2011temporal},
can one use such information to infer more details about weather-mood correlation?

However, Twitter data exists in an extremely noisy structure.
In particular, 
extracting tweets that explicitly express statements of mood is not an easy task. 
Researchers usually turn to rough analysis by determining mood or sentiment
considering simply the presence or absence of sentiment terms, which can be problematic in some tasks.
For example, 
`happy' in `happy new year' is not considered to be a reflection of personal happiness but rather a conventional phrase;
 ``I am not happy today" will be interpreted as positive when applying keyword matching.
Additionally, the change of individual mood can be directly induced by non-weather related events such as political events, sport events, natural disasters.
For example, {\it Haiti Earthquake} or {\it Michael Jackson's death} arouse a widespread sorrow. 
Ignoring these highly influential factors can be damaging.

To address the aforementioned issues, we proposed a sophisticated machine learning pipeline approach to select explicit mood expressing indicators and get rid of the nuisance of public event factor.
We further conduct a detailed investigation of correlation between multiple dimensional structure of human mood and weather factors.
We make theoretical contributions, some of which confirm longstanding hypotheses: for example, people tend to be more furious
at high temperatures. As another example, snow may aggregate depression and is responsible for increased suicide rate \cite{rind1996effect} or aggregation of Seasonal Affective Disorder \cite{leppamaki2002bright}.

It is still worth noting that as in the cases of many researches in the social science literature, there are no gold standards to test our findings or the solidness of the proposed approach. 
Additionally, due to the extremely complicated circumstances involved in the causality of mood states, any analysis in practice will inevitably overlook a myriad of factors that may directly or indirectly intervene in the relation analysis (see Section 6). Nevertheless, our approach still offers obvious advantages over existing methodology regarding this problem, and in some sense explains, verifies or confirms some of the existing hypotheses. 
We hope that the proposed framework could shed light on a longstanding debate in the literature about weather-mood relationship.

The rest of this paper is organized as follows: 
We go over related work in Section 2 and describe our data sources in Section 3. Related processing algorithms are described in Section 4. We present our findings and analysis in Section 5, followed by a discussion and conclusion in Section 6. 

\section{Related Work}
In this section, we briefly go through the related work.\\~\\
{\bf Mood Weather Relationship in Psychology and Physiology}\\~~\\
It has long been believed that there should be some sort of correlation between weather and individual mood state. 
In psychological literature, researches on this topic can be traced back to early 1950s when Nowis \cite{nowlis1956description}  tried to explore the external factors that affect individuals' emotions. 
While correlational analysis between mood and weather has been conducted in multiple psychological researches \cite{clark1988mood, cunningham1979weather,goldstein1972weather,howarth1984multidimensional,persinger1975lag,rind1996effect,watson2000mood}, conclusions are amazingly diverse. 
Low levels of humidity \cite{sanders1982relationships}, high levels of sunlight \cite{cunningham1979weather,parrott1990mood,schwarz1991evaluating}, high barometric pressure \cite{goldstein1972weather} and high temperatures \cite{cunningham1979weather, howarth1984multidimensional} have been associated with high mood, but high temperature has been associated with low mood in some other work \cite{goldstein1972weather,howarth1984multidimensional}.
Meanwhile, some studies traced no direct correlations between mood and any type of weather variables \cite{clark1988mood,watson2000mood}: Watson and Lee \cite{watson2000mood} traced the daily mood reports of 498 students in Dallas, Texas and found no significant correlations between mood (measured by self-report using the Positive and Negative Affect Scale) and any of the assessed weather variables, including sunshine, barometric pressure, temperature, and precipitation.

Despite the on-going debates in psychology, most physiologists believe that the weather indeed has significant influence on psychological processes based on strong physiological evidence. For example, Lambert et al. \cite{lambert2002effect} spotted rising and dipping production of serotonin, a neurotransmitter in human central nervous system, which is generally thought to be a contributor to feelings of well-being and happiness, as sunlight increased or decreased.
Another line of evidence is provided by population-wide behavior studies in social psychology.
A good example is seasonal affective disorder (SAD), which refers to the mood disorder in which people who have normal mental health throughout most of the year experience depressive symptoms in the winter\footnote{\url{http://en.wikipedia.org/wiki/Seasonal_affective_disorder}}. Researchers suggest that exposure to sunlight would immediately improve mood and diminishes SAD \cite{kripke1998light,lambert2002effect,leppamaki2002bright,stain1998platelet} and that rainy or cloudy days can aggravate the symptoms \cite{leppamaki2002bright}. 
Additionally, reliable and significant associations are found between weather and mood-related social behaviors. 
For example, low barometric pressure \cite{digon1966suicides} and more snow are evidently linked with higher suicide rates  \cite{rind1996effect} and high temperature is associated with violent behavior \cite{anderson2001heat,baron1978ambient}.

~\par

\noindent{\bf Mood/Opinion Tracking based on Social Media}\\~~\\
There has been significant progress
in extracting indicators of public mood
in a variety types of online social media, such as blogs \cite{dodds2010measuring,gilbert2010widespread,li2014timeline,liu2007arsa,mishne2006capturing}, Twitter \cite{pak2010twitter} or Facebook.
Facebook introduced Gross National Happiness (GNH) that records the positive and negative words used in status updated by Facebook users in an attempt to estimate the aggregated mood and well-being of the 
Facebook population.
GNH was built upon identify mood/sentiment words appearing in status updates based on LIWC dictionary and results show a clear a weekly cycle and increases on national holidays \cite{dodds2011temporal}. Based on GNH, Kramer \shortcite{kramer2010unobtrusive} found happier individuals use more positive words and fewer negative words in their status updates on Facebook.
Real-time mood tracking website such as Pulse of Nation\footnote{\url{http://www.ccs.neu.edu/home/amislove/twittermood/}} are established to keep trace of public mood based on mood signals from online social media.
Mitchell et al. \shortcite{mitchell2013geography}  investigate how geographic place correlates with and potentially influences societal levels of happiness from Twitter data.
In addition to the merit of great multitude of available messages, another good property that Twitter data exhibits, overlooked by most researchers, is the avoidance of {\it volunteer effect}, which indicates the incident of participation in the survey will influence the psychological states of volunteers. 

As Twitter mood indicators have been demonstrated as both robust and sensitive \cite{dodds2010measuring,dodds2011temporal}, researchers are exploring ways to predict a variety of real-world outcomes based on indicators, such as economics \cite{gilbert2010widespread,rao2012tweetsmart,rao2012using,zhang2011predicting}, stock market\cite{bollen2011twitter,chung2011predicting} or political events \cite{hopkins2010method,o2010tweets,tumasjan2010predicting,larsson2012studying} or disease \cite{lampos2010flu,li2013early}.  
Lindsay \cite{lindsay2008predicting} built a sentiment classifier with Facebook data set and demonstrates the significant correlation between microblog sentiments and polls during 2008 presidential election.
O'Connor et al. \cite{o2010tweets} tried to predict political events based on user-mood analysis on twitter.
\begin{figure*}
\includegraphics[width=1\textwidth]{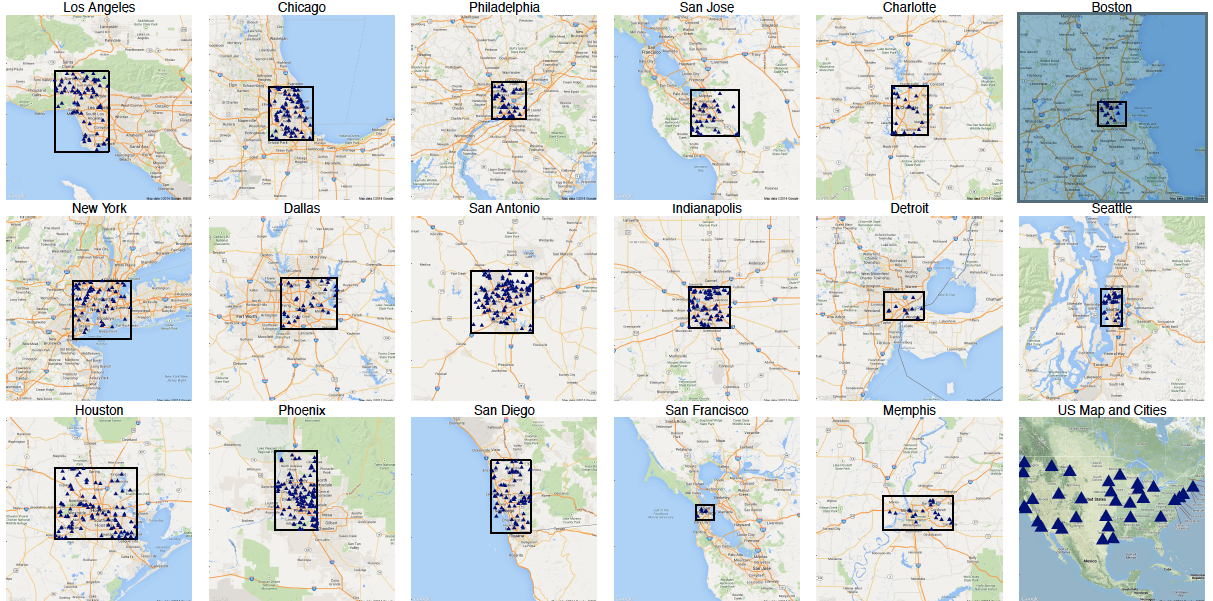}
\centering
\caption{\label{fig:geoloc}Illustration of NOAA station data: Geographic locations of survey urban areas (blue upper triangles), boundaries (blue rectangles) and climate stations (blue dots).}
\end{figure*}

\section{Data Sources}
We begin with describing the sources of data used in this study: Twitter Corpus, Meteorological dataset and opinion vocabulary sources.
\subsection{ Twitter Corpus}
We examine a corpus of geo-tagged tweets gathered from 32 urban areas in the United States. 
The corpus is a CMU subset of garden hose, which represents roughly $10\%$ of all Twitter postings in 2010 and 2011.
For the present study, we focus on the approximately $1\%$ or $2\%$ tweets that are geo-tagged. 
Urban areas are defined by the 2010 United States Census Bureau's MAF/TIGER (Master Address File/Topologically Integrated Geographic Encoding and Referencing) database\footnote{\url{http://www.census.gov/geo/www/tiger/tgrshp2010/tgrshp2010}}.

\subsection{Meteorological Data from NOAA}
{\it NOAA}, short for National Oceanic and Atmospheric Administration\footnote{\url{http://www.noaa.gov/}}, is a scientific agency within the United States Department of Commerce focused on the conditions of the oceans and the atmosphere\footnote{\url{http://en.wikipedia.org/wiki/National_Oceanic_and_Atmospheric_Administration}}. 
Two data sources of NOAA constitute our meteorological dataset: 
USRCN (U.S. Climate Reference Network)\footnote{https://www.ncdc.noaa.gov/crn/} and GHCN (Global Historical Climatology Network)\footnote{http://www.ncdc.noaa.gov/oa/climate/ghcn-daily/}, both of which belong to 
NOAA's National Climate Data Center\footnote{http://www.ncdc.noaa.gov/}.
Meteorological observation station is mapped to correspondent Urban areas 
based on geographic bounds geocoded using R package \textit{ggmap} \cite{ggmap},
as shown in Figure~\ref{fig:geoloc}. 
Meteorological factors we explore include 
\begin{itemize}
\item Average Temperature (tAVE): $^\circ \rm C$
\item Daily Temperature Change (tCHANGE): $^\circ \rm C$
\item Daily Precipitation (PRCP): mm
\item Daily Snow Depth (SNWD): mm
\item Average Daily Wind Speed (AWND): meter/second
\item Total Solar Energy Received (TSER): MJ/meter$^2$
\item Hail, Thunder, Smoke, Fog, Frost: presence or absence
\end{itemize}

Daily Temperature Difference (tCHANGE) 
is obtained by taking the average of temperature difference collected at the same time from two consecutive days.

\subsection{OpinionFind and Profile of Mood States}
OpinionFinder \cite{wilson2005recognizing} is a sentence-level sentiment analysis software package and has been widely used to analyze emotional content in large collections of tweets \cite{benson2011event,o2010tweets}. We use OpinionFind's subjective lexicons established from previous work \cite{riloff2003learning,riloff2003learning1} containing1,600 positive and 1,200 negative labeled terms.

The fact that OpinionFinder only distinguishes between positive and negative sentiment ignores the rich dimensional state of human mood \cite{bollen2011twitter}. We additionally turn to {\it Profile of Mood States} (POMS) \cite{norcross1984factor,pollock1979profile,shacham1983shortened}, a psychological rating scale used to assess transient, distinct mood states.
We explore 3 dimensions of mood state based on POMS
, {\it Anger-Hostility}, {\it Fatigue-Inertia} and {\it Depression-Dejection} and a manually added {\it Sleepiness-Freshness} dimension. In practice, we adopted the approach introduced in Bollen et al. 's work \shortcite{bollen2011twitter}  
by expanding the original term list provided by POMS questionnaire from word co-occurrence analysis in a collection of 4 and 5-grams computed from Google \cite{bergsma2009web}. Then each tweet term that matches expanding terms is mapped back to original POMS terms. The enlarged lexicon list contains 162 terms and allows us to capture a larger varieties of mood related tweets.

\section{Selecting Mood State Indicators}
In the section, we describe preprocessing algorithms involved in our approach, which aims at electing out tweets that explicitly include individual mood expression. 
\subsection{Identification of Public Event-related Tweets}

As we discussed before, we wish to the largest extent, filtering out sentiment indicators induced by public events. This is necessary since public can be up about the release of popular movie or sports game and down about nature disasters, 
or the death of public figures.
See the following examples where [ ]  denote mood indicators:  
\begin{itemize}
\item $\#$Haiti on my mind. Throughout all our [pain] and [sorrow] from the earthquake we still stand strong.
\item It's a [sad] day for pop. News of Michael Jackson's death has come as a [tragic] shock.
\end{itemize}
The first tweet corresponds to the public event {\it Haiti Earthquake} and the second to {\it The death of Michael Jackson}, both of which aroused widespread sorrow. 
To get rid of mood indicators induced by public events, 
we first employed Alan Ritter's Twitter pipeline system \cite{ritter2012open} which aims at extracting open domain public events from Twitter.
The system first identifies name entities and then uses a CRF-based approach
for event phrase identification. Name entities and event mentions are next fit into a LDA topic model for event clustering. Event related clusters mined from LDA are manually identified, where 52 clusters are found to correspond
to coherent event types referring to significant events such as Movie, Disaster or Sports. 
In the aforementioned examples, $Haiti$ and $Michael Jackson$ will be identified as name entities while ``earthquake" and ``death"
as event mentions, which will all be later classified as certain type of public event from LDA topic model.
Part of examples are shown in Table \ref{tab2}. Postings containing event-related mentions are discarded. 

\begin{table}
\centering
\small
\begin{tabular}{|c|c|}\hline
manual label&top words\\\hline
sports&NFL, Lebron James, Tigers, Brees\\\hline
disaster&Haiti Earthquake, earthquake, died, suffer\\\hline
movies&movie, netflix, black swan, inception\\\hline
politics& Obama, debate, debates, america, president\\\hline
war& Libia, war, Gaddafi, Syria, bomb\\\hline
death& Micheal Jackson, sad news, died\\
&Whitney Houston\\\hline
TV&	Idol, Idol season, Big Bang Theory\\\hline
music&Justin Bieber, Feat, taylor swift\\\hline
\end{tabular}
\caption{Example event types, reprinted from Ritter et al. 's system \cite{ritter2012open}.}\label{tab2}
\end{table}

\subsection{Identifying Mood State Related Tweets}
To note, we first narrow tweet candidates to sentiment indicator containing tweets 
by matching bag of words from OpinionFind and Profile of Mood States.
As we mentioned, not all tweets containing mood indicators explicitly indicate the mood state of users.
Bollen et al. \shortcite{bollen2011twitter} perform simple bag of words matching such as ``I feel", ``I am feeling" for identifying 
mood related tweets which can easily violated by expressions such as ``I feel like having McDonald for lunch ".
Towards a better resolution, we train a Max-Ent classifier to distinguish between positive, negative and non-mood-related tweets. Features we explore include:
\begin{itemize}
\item Tweet-level: token identity from the posting, correspondent POS.
\item Sentiment indicator: token identity, POS
\item Window feature: left and right context words within 3 window of words and the correspondent POS.
\end{itemize}

Part-of-speech tags are assigned based on Owoputi et al. 's tweet POS system \cite{owoputi2013improved}.
We manually labeled 600 tweets that contain mood indicators as positive, negative or non-mood-related and
performed a 5-fold cross-validation where
model parameters are selected for each test fold based on experiments on the training folds.
The F-1 scores regarding positive, negative and non-sentiment are 0.782, 0.806 and 0.871 respectively.
We further group mood-related tweets into different mood dimensions by matching keywords from the aforementioned expanded POMS term list.   
\subsection{Opinion Estimation}
We derive day-to-day sentiment scores by counting positive and negative messages. 
As in most previous researches \cite{o2010tweets,zhang2011predicting}, 
we increase the score of either negative or positive tweets identified previously.
We adopt the tradition where  
the mood rating score $x_t$ on day $t$
is defined
as the ratio of \textsc{positive} versus overall, counting from that day's messages after public-topic filtering and Max-Ent classifier selection:
\begin{equation}
x_t=\frac{count_t(positive)}{count_t(negative)+count_t(positive)}
\end{equation}

\section{Correlation Analysis}
In this section, we present our results in detail. We first present our correlation analysis methodology, then our binary mood analysis (positive/negative) and finally move on to different dimensions of mood states.
\subsection{Statistical Analyzer}
We applied a Generalized Mixed Model (GMM) \cite{gam,wood2011} to capture the non-linear relationship between mood  and climate factors. As the detail is quite standard

where we assume the mood rating score $x_t$
follows a Poisson process depending on a number of climate covariates. 
Concretely,
through a link log function $h(\cdot)$, 
$x_t$ is
 explained by additive non-linear functions of climate variables, their higher-order correlation part and self-correlated time trends.
 Each non-linear function is non-parametrically decomposed to a set of basis Cubic Spline function\footnote{\url{http://en.wikipedia.org/wiki/Spline_(mathematics)}} in this paper. 

The adopted GMM models takes into consideration the following factors (1) time autocorrelation (2) the inter-correlation between regression variables (3) external information (e.g. city-specific factors, weekday factors) by adding additional variables.
Inference is implemented by maximizing penalized likelihood \cite{wood2000} where the penalized parameters are tuned based on generalized cross-validation method \cite{wood2004}. The knots of Spline are penalized to avoid over-fitting.

\subsection{Positive/Negative Analysis}

\begin{figure*}
\includegraphics[width=3.5in]{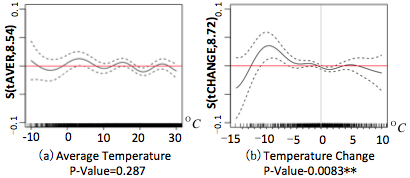}
\includegraphics[width=3.5in]{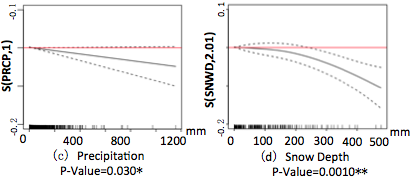}
\includegraphics[width=3.6in]{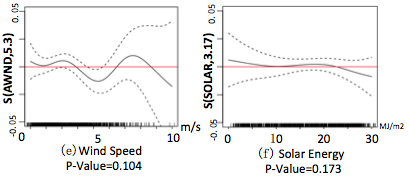}
\includegraphics[width=3.3in]{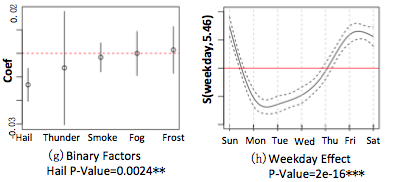}
\caption{Positive/Negative mode analysis regarding multiple meteorological factors. Red solid line corresponds to 0 line. Black dotted lines correspond to boundary of confidence interval. Black solid line corresponds to regression curve. y-axis corresponds to smooth regression value from GAM model. 
Positive value of smooth regression means positive contribution to up-mood state while negative value means the opposite.
Label for y-axis corresponds to {\bf S(meteorological factor, degree of freedom)}}.
\label{fig3}
\end{figure*}

We report our findings regarding positive/negative mood state in this subsection.
For each meteorological factor, we report regression results along with the correspondent $95\%$ confidence interval in Figure 3. 
y-axis {\bf S} corresponds to the smooth regression coefficient. Positive value of {\bf S} indicates positive contribution to up-mood state while negative value indicates the opposite.
Notably, {\bf zero} point not being included in the confidence interval indicates significant confidence.
~\par~\par
\noindent {\bf Daily Average Temperature (tAVER):} 
As we can observe from Figure 3(a), temperature does not make significant contribution to mood state with P-Value=0.287. 
The regression curve fluctuates irregularly and
0 point is mostly included in the confidence interval.
Our conclusion with temperature is in agreement with some of early findings \cite{clark1988mood,watson2000mood}. 
~\par~\par

\noindent {\bf Daily Temperature Change (tCHANGE):} 
While mood state is not sensitive to temperature, it is significantly sensitive to temperature change (P-Value=0.0083), as shown in Figure 3(b). 
According to our findings, people tend to be happier as temperature becomes cooler
but feel uncomfortable with drastic temperature decrease. 
People are a little bit negative towards small temperature increase, but their responses to 
great temperature increase are unclear.

~\par~\par
\noindent {\bf Daily Precipitation (PRCP):} 
Our findings with precipitation are largely in accord with expectation as it has negative effect on individuals' mood, 
as shown in Figure \ref{fig3}(c).  
The negative influence becomes greater as precipitation increases, which a little bit deviates from common sense, as we think 
people tend to be more annoyed when drizzling. 
~\par~\par

\noindent {\bf Daily Snow Depth (SNWD)}
Mood is significantly affected by the depth of snow (P-value=0.0010). More snow leads to negative mood state, as shown Figure \ref{fig3}(b). 
Our findings 
to some extent explain early observations such as snow leading to higher suicide rate \cite{rind1996effect} or aggregating SAD symptoms \cite{harmatz2000seasonal}. 
~\par~\par
\noindent {\bf Average Daily Wind Speed (AWND) }:
While great wind power is associated with negative mood in previous researches \cite{denissen2008effects}, 
according to our results, wind does not significantly contribute to mood state (P-value=0.104).
~\par~\par
~\par~\par

\noindent {\bf  Total Solar Energy Received (TSER)}: 
TSER can be interpreted as the combined measurement of hours of daylight and sunlight strength.
Different from our expectation, mood is not significantly affected by TSER.
As people's mood state is not sensitive to temperature, it is not supervising that less responses  are found to Solar Energy. 

~\par~\par
\noindent{\bf Hail, Thunder, Smoke, Fog, Frost}: Among five binary-valued weather variables, only Hail makes significant contribution (P-value = 0.0024), as 
is significantly associated with negative mood. The influence from remaining variables is unclear.
~\par~\par
\noindent{\bf Side Variable 1, Urban Area:} We include Urban Area in our analysis as we wish to exclude undesirable influences that are city-specific, usually caused by difference of
living standard, population composition, employment rate or degree of industrialization  \cite{mitchell2013geography}.
In accord with previous findings \cite{mitchell2013geography},
influence from Urban Area is extremely significant (P-Value=2e-15). As urban happiness is not the theme of this paper, the visualization is excluded for brevity.
~\par~\par
\noindent{\bf Side Variable 2, Weekday Effect:} Individual's mood exhibits a clear fluctuating pattern within the week, 
up on weekends and down on weekdays as shown in Figure \ref{fig3}(h). Similar patterns are
spotted in many previous researches \cite{dodds2011temporal,kramer2010unobtrusive}. 

\begin{figure*}[!ht]
\centering
\includegraphics[width=3.3in]{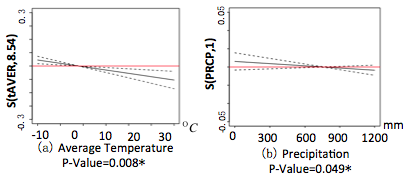}
\includegraphics[width=3.1in]{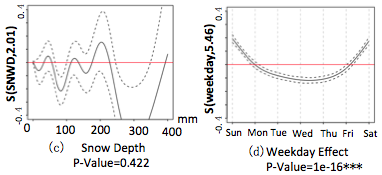}
\caption{Correlation Analysis with {\bf Hostility-Anger}}.
\end{figure*}

\begin{figure*}[!ht]
\vspace{-0.5cm}
\centering
\includegraphics[width=3.3in]{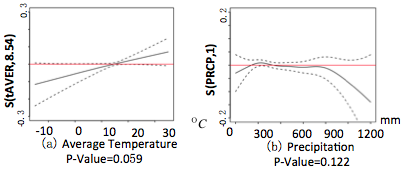}
\includegraphics[width=3.1in]{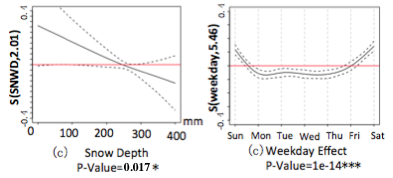}
\caption{Correlation Analysis with {\bf Depression-Dejection}}.
\end{figure*}

\begin{figure*}[!ht]
\vspace{-0.5cm}
\centering
\includegraphics[width=3.3in]{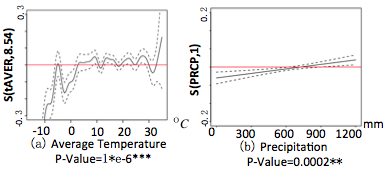}
\includegraphics[width=3.2in]{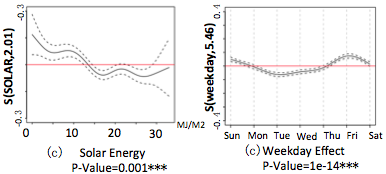}
\caption{Correlation Analysis with {\bf Fatigue-Inertia}}.
\end{figure*}

\begin{figure*}[!ht]
\vspace{-0.5cm}
\centering
\includegraphics[width=3.3in]{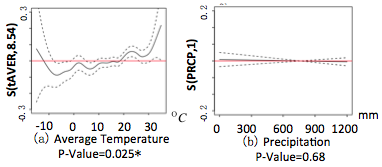}
\includegraphics[width=3.2in]{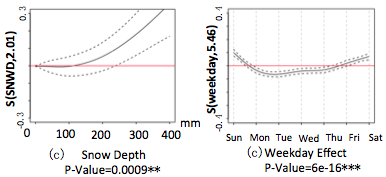}
\caption{Correlation Analysis with {\bf Sleepiness-Freshness}}.
\end{figure*}

\subsection{Dimensional Analysis}
In this section, we explore the influence of weather factors on each of the 4 mood dimensions, 
namely, Hostility-Anger, Depression-Dejection, Fatigue-Inertia and Sleepiness-Freshness.
We report our results in Figure 4. We focus our attention on 4 main meteorological factors, Temperature, Snow, 
Precipitation and Solar Energy, along with Weekday Effect. Note that in Figure 4, higher regression value indicates more contribution 
to positive mood dimension, in other words, less angry, less depressed, less tired and less sleepy.

~\par~\par
\noindent{\bf Hostility-Anger}: 
While influence from temperature on total positive/negative mood state is trivial and vague, 
it is significantly linked with anger: the hotter, the angrier, 
as we can observe from Figure 4(a). Precipitation has a small tendency to provoke anger, but not obvious (P-Value=0.049).  
The contribution of Snow is random as shown in Figure 4(b).
Weekday Effect again makes significant contribution.
~\par~\par
\noindent{\bf Depression-Dejection}:  Snow Depth lead to Depression (P-Value = 0.017), as expected (see Figure 5 (c)). The influence from other meteorological factors are unclear. Higher average temperature tends to alleviate depression and precipitation may aggregate depression, according to the tread of the curves (see Figure 5 (a)(b)). But their correspondent P-Values indicate the lack of confidence. 
~\par~\par
\noindent{\bf Fatigue-Inertia}: While the curve of temperature fluctuates, as shown in Figure 6(a), 
we can draw the general conclusion that high temperature leads to tiredness, as the value of smooth regression value mostly takes negative value at temperature below 0 but positive at temperature above 0.
Higher solar energy increases the feeling of tiredness, as clearly demonstrated in Figure 6(c), which is in accord with 
early findings from Jaap et al. 's work \shortcite{denissen2008effects}.
Precipitation alleviates tiredness, shown in Figure 6(b).

Another interesting thing about Tiredness is, different from weekday trends of all other factors (See Figures 4(d), 5(d), 7(d)) where peaks all appear on weekends, indicating people tend to be the least angry, the least depressed and the least sleepy on weekends, the curve of Fatigue-Inertia arrives at its peak on Friday. 
~\par~\par
\noindent{\bf Sleepiness-Freshness}: Common senses tell us people tend to be fresh at low temperature, which is 
not in accord with our findings. Evidence from Figure 7(a) indicates cool temperature is lined with 
sleepiness and people tend to be fresher and fresher as temperature increases.
Precipitation has not significant influence on this dimension but snow significantly contributes to freshness. 

\section{Conclusion and Discussion}
In this paper, we propose a framework that harnesses Twitter data, trying to answer the long-lasting question about mood-weather correlation. Our sophisticated pipeline to the largest extent gets rid of undesirable influence. Our approach explains, confirms or contradicts some of existing hypotheses or findings. 

It is worth noting that, the proposed approach comes with the following problems in nature: 

\noindent {\bf Data Level}:
At data level, while Twitter data has verified a series of long-standing hypotheses in social sciences, 
which in turn demonstrates
the robustness and sensitiveness of this source of data, 
there is no generally accepted or rigorously proved theory on how straightforward 
the 
relationship between Twitter status updates and general mood or well-being of users would be.
People may try to disguise their real feelings and
present their Twitter friends a contrived image. As such relationship is infeasible to evaluate, there is
no guarantee that the methodology is well grounded.
Scarce as it is, there has already been criticism for the validity of aggregate level mood/sentiment analysis based on data source from online social media \cite{wang2012can}. 
Wang et al. \cite{wang2012can} proposed two plausible explanations for the possible discrepancy between status updates on social media and real mood of users (1) users disguising their feelings, as just mentioned above (2) failure of current machine learning analysis tools to correctly decipher individual's mood or sentiment.
To be specific for our task, Twitter users may not be a representative sample of the real-world and the geo-tagged tweets can come with a significant collection bias. There is also the bias of users deciding to report their moods. Additionally, people's mood is affected by all sorts of situations unrelated to the weather. Even though we hope the {\it aggregated analysis} will help leverages the all these unrelated effects, it is unclear how it work.

\noindent {\bf Statistical Level}:
At statistical level, the general tricky problems involved in 
most statistics based analysis also confront us, namely \textsc{Confounding}\footnote{\url{http://en.wikipedia.org/wiki/Confounding}} and
\textsc{multicollinearity}.
\textsc{Confounding} points to a latent variable in a statistical model that correlates (directly or inversely) with regression variable.
The failure to account for a confounding factor usually 
leads to a spurious relationship or correlation. For example, the presence of ozone (O$_3$) is generally linked with a fresh head and up mood \cite{tom1981influence}. However, O$_3$ is usually associated with emissions from automobiles and photochemical air pollution in real life. As the latter can be strongly linked with down-mood factors such as stress and depression \cite{clougherty2010chronic}, relation analysis between mood and ozone without considering air pollution effect will lead to erroneous conclusions.
For another example, 
snow leads to down mood, as we find in the above analysis.
However, it is unclear whether the widespread down mood is caused by snow itself or other snow induced factors, such traffic jams, traffic accidents or other unseen factors. As enumerating all related factors in real world and collecting correspondent data are infeasible, it is hard to say which is primarily responsible for the down mood, snow or snow-induced factors.

\textsc{multicollinearity} refers to statistical phenomenon in which two or more predictor variables in a multiple regression model are highly correlated\footnote{\url{http://en.wikipedia.org/wiki/Multicollinearity}}. 
In the cases where data can manifest multicollinearity itself, we try to consider its effect by adding high-order interaction part. 
Unfortunately, multicollinearity can exist in a more implicit way and is therefore hard to identify or evaluate.
For example,
while the significant influence from Urban Area are mostly cause by city-level features such as living standard, population composition, employment rate or degree of industrialization\cite{mitchell2013geography}, 
weather itself can also be city-specific. For instance, cities resided on plateau (i.e., Denver) usually have a lower value of barometer pressure. More urban areas being considered can alleviate but can't dispel the nuisance. 
Of course we can perform analysis city by city and then obtain a global estimation from a mixture model\footnote{\url{http://en.wikipedia.org/wiki/Mixture_model}}, but it will additionally induce problems such as 
locality or complex inference. The value of Deviance Explained\footnote{\url{http://en.wikipedia.org/wiki/Deviance_(statistics)}} 
confirm the benefit of the adopted analysis model over mixture model ($44.2\%$ versus $30.8\%$).
In summary, the impossibility of ruling out the influence of 
\textsc{Confounding} and
\textsc{multicollinearity} can lead to biased conclusions.\\

In this study, we can only acknowledge that we are aware of these pitfalls and try to come up with a more accurate and sophisticated  machine learning system to avoid them.
Despite of disadvantages discussed above, 
the proposed approach still offers significant benefits over methodologies adopted in standard psychology studies and we hope it would offer a new prospective toward the understanding for mood-weather relationship and 
could shed light on this long-standing problem.\\

\noindent {\bf Acknowledgement}\\
Data used in this work is extracted from CMU Twitter Warehouse maintained by Brendan O'Connor, to whom we wish to express our gratitude. 
We also want to thank CIKM reviews for useful suggestions and comment. 


\bibliographystyle{abbrv}
\bibliography{mybiblio}  


\end{document}